%% file: MLPH_2021-assisted-living-in-the-united-states.tex
\documentclass{article}
\usepackage[nonatbib,final]{neurips_2021}
\usepackage{verbatim}

\input{preamble/preamble}
\input{preamble/preamble_acronyms}
\input{preamble/preamble_biblatex}
\newcommand{\numberofalfs}{44,638}

\title{
Assisted Living in the United States: an Open Dataset}

\author{%
  Anton Stengel\thanks{These authors contributed equally to this work.}\\
  Princeton University\\
  \href{mailto:astengel@princeton.edu}{astengel@princeton.edu} \\
  \And
  Jaan Altosaar\footnotemark[1]\\
  One Fact Foundation\\
  \href{mailto:jaan@onefact.org}{jaan@onefact.org}\\
  \AND 
  Rebecca Dittrich\\
  ATI Advisory\thanks{Work conducted while affiliated with ATI Advisory.}\\
  \href{mailto:rebecca@atiadvisory.com}{rebecca@atiadvisory.com}\\
  \And
  Noémie Elhadad\\
  Columbia University\\
  \href{mailto:noemie.elhadad@columbia.edu}{noemie.elhadad@columbia.edu}\\
}

\begin{document}

\maketitle

\begin{abstract}

An \gls{alf} is a place where someone can live, have access to social supports such as transportation, and receive assistance with the activities of daily living such as toileting and dressing. Despite the important role of \glspl{alf}, they are not required to be certified with Medicare and there is no public national database of these facilities. 
We present the first public dataset of assisted living facilities in the United States, covering all 50 states and DC with \numberofalfs~facilities and over 1.2 million beds. This dataset can help provide answers to existing public health questions as well as help those in need find a facility. The dataset was validated by replicating the results of a nationwide study of \glspl{alf} that uses closed data \cite{cornell_developments_2020}, where the prevalence of \glspl{alf} is assessed with respect to county-level socioeconomic variables related to health disparity such as race, disability, and income. To showcase the value of this dataset, we also propose a novel metric to assess access to community-based care. We calculate the average distance an individual in need must travel in order to reach an \gls{alf}. The dataset and all relevant code are available at \href{https://github.com/onefact/assisted-living}{github.com/onefact/assisted-living}.
\end{abstract}

\section{Introduction}
\label{sec:intro}

The coronavirus pandemic has highlighted several concerns about the lack of access to alternative options for housing an aging population in the United States. Deficiencies in nursing facilities exposed by the pandemic have motivated policy conversations around the need to support opportunities for older adults to age at home or in the community. 

One of the largest community-based care options in the United States is the system of \glspl{alf}. For comparison, there are around 15,000 licensed nursing homes in the United States \citep{harris-kojetin_long-term_2019}, while the number of \glspl{alf} is over 40,000 \cite{cornell_developments_2020}, with both systems housing over one million people each. As over half of Americans turning 65 today are projected to require some form of long-term assistance due to disability \citep{favreault_long-term_2021}, the need for greater access to long-term services and supports will become even more important.

\input{table_replication}

An \gls{alf} is a place where someone can live, have access to social supports such as transportation, and receive assistance with activities of daily living.
For example, an individual with serious mental illness who lives in an \gls{alf} may need staff assistance with taking medications, while an elderly person living in an \gls{alf} may require help with activities of daily living, such as getting dressed or eating.

But while over one million people live and receive care in \glspl{alf} in the United States, there is no public dataset with which to quantify the number of facilities, how they are licensed, and where they are located, since they are regulated on a state-by-state basis. Further, it is estimated that there are a significant number of unlicensed facilities in some states \citep{greene_understanding_2015}; the possibility of abuse of the elderly or people with serious mental illness at such unlicensed locations further highlights the lack of transparency and data on community-based care. 

We build the first public dataset of \glspl{alf}, validate the dataset by replicating recent work that studies access to community-based care, and propose a new metric to understand community-based care and health equity at a national level. By documenting the data collection process, we highlight the difficulty of data access. This illustrates some of the structural barriers to what should be public and easily-accessible data: these barriers pose a serious problem to the public, policymakers, and researchers. We hope that this dataset will enable increased transparency and accountability of \gls{alf} licensing, and enable the development of machine learning methods to answer public health questions \cite{mhasawade_machine_2021}, such as how best to expand access to community-based care in the United States and other countries.

\section{Building an Open Dataset of Assisted Living Facilities}
\label{sec:building-dataset}
Unlike nursing homes, \glspl{alf} are not federally regulated. Each state licenses different forms of \glspl{alf}, under varying names and varying regulations. For example, states may license these facilities as \glspl{alf}, Residential Care Facilities, Housing with Services, Homes for the Aged, Residential Health Care Facilities, Shared Housing, Personal Care Homes, among other license types. 


We follow the National Center for Assisted Living's 2019 regulatory review of assisted living 
to define an \gls{alf} \cite{noauthor_assisted_2019}. This review delineates the regulatory requirements for \glspl{alf} based on state-by-state license types and other criteria. We verified this definition with how the federal government defines \glspl{alf} (as residential care communities) in its National Post-Acute and Long-Term Care Study \cite{noauthor_national_2018}.

\paragraph{Data Collection.} All \glspl{alf} collected for this dataset came from state licensing agencies in the summer of 2021, but with significant differences in the ease, transparency, or reproducibility of data access and collection. Of the 50 states and the District of Columbia, 20 states did not have a satisfactory dataset of \glspl{alf} online, necessitating a byzantine series of steps to acquire data. 

As an illustrative anecdote of the structural barriers to data access, we describe the collection process for Arkansas. For this state, the data was not available on a state website and thus required emailing a state licensing agency. This agency was only able to provide half of the dataset of \glspl{alf}; the other half is licensed by a different agency in the state. This resulted in a referral to the second department, which had a policy not to share licensed facilities and meant we had to submit a Freedom of Information Act request. The first such request was denied and had to be resubmitted by a resident of the state on our behalf.

For states with data available online, half a dozen states only provided data directly in their webpage and half a dozen only provided data as a PDF. Data directly on webpages was parsed to extract information into a \gls{csv}. For data available as a PDF, we used optical character recognition to convert the PDF into text and then extracted the textual information into a \gls{csv}. For the remaining states, the data was available as a \gls{csv}.

For the rest of the states, licensing agencies were contacted directly. Some states provided data as a \gls{csv} and some states only released data as a PDF. Data had to be purchased from one state for \$15 (and turned out to have the highest level of missing attributes of any state). Four states, such as Arkansas, refused to provide a directory of licensed \glspl{alf}. For these states, \gls{foia} requests were submitted. Two requests were approved, and two were denied, and we had to contact residents of the state who then successfully submitted the \gls{foia} requests on our behalf. After the data was collected, it was cleaned and pre-processed as described in \Cref{sec:data-cleaning}.

\paragraph{Linking assisted living facilities to American Community Survey data.} We augmented the \gls{alf} dataset with relevant societal metrics. Each \gls{alf} is connected to county information on racial demographics and socioeconomic statistics retrieved from the American Community Survey (ACS) 2015-2019 longitudinal study \citep{noauthor_american_nodate}. Additionally, we associate each \gls{alf} with the need-based metric for community care which is described in Section \ref{sec:need-metric}.

Details about the dataset are described in \Cref{sec:datasheets} using the dataset documentation framework from \citet{gebru_datasheets_2020}.

\section{Validation}
\label{sec:validation}

To highlight the utility of an open dataset for answering public health questions with computational methods, we first undertake a conceptual replication of the analysis of \citet{cornell_developments_2020} by assessing the prevalence of \glspl{alf} with respect to socioeconomic characteristics.  Specifically, we replicate Table 1 in \citet{cornell_developments_2020} in the present \Cref{table:replication}. 
While the numbers are overall quite similar, one difference in the replication is that the number of counties with no \glspl{alf} in our conceptual replication study is higher than reported in \cite{cornell_developments_2020}. One potential reason for this is that \citet{cornell_developments_2020} focus on county-level capacity information which is easier to access compared to assessing individual facilities as we do in this dataset. This also highlights the need for transparent data collection methods (\citet{cornell_developments_2020} do not specify how data on \glspl{alf} was collected). Perhaps the method we describe in \Cref{sec:building-dataset} of reaching out to state licensing agencies leads to less capacity information than other (potentially non-public) methods of data access. Despite this difference, the statistics between \citet{cornell_developments_2020} and our study are comparable, validating the overall aims of the dataset to enable assessment of access to community-based care and correlation to health equity measures.

\begin{figure}
    \centering
    \includegraphics[width=1\linewidth,trim={0 0 0 2cm},clip]{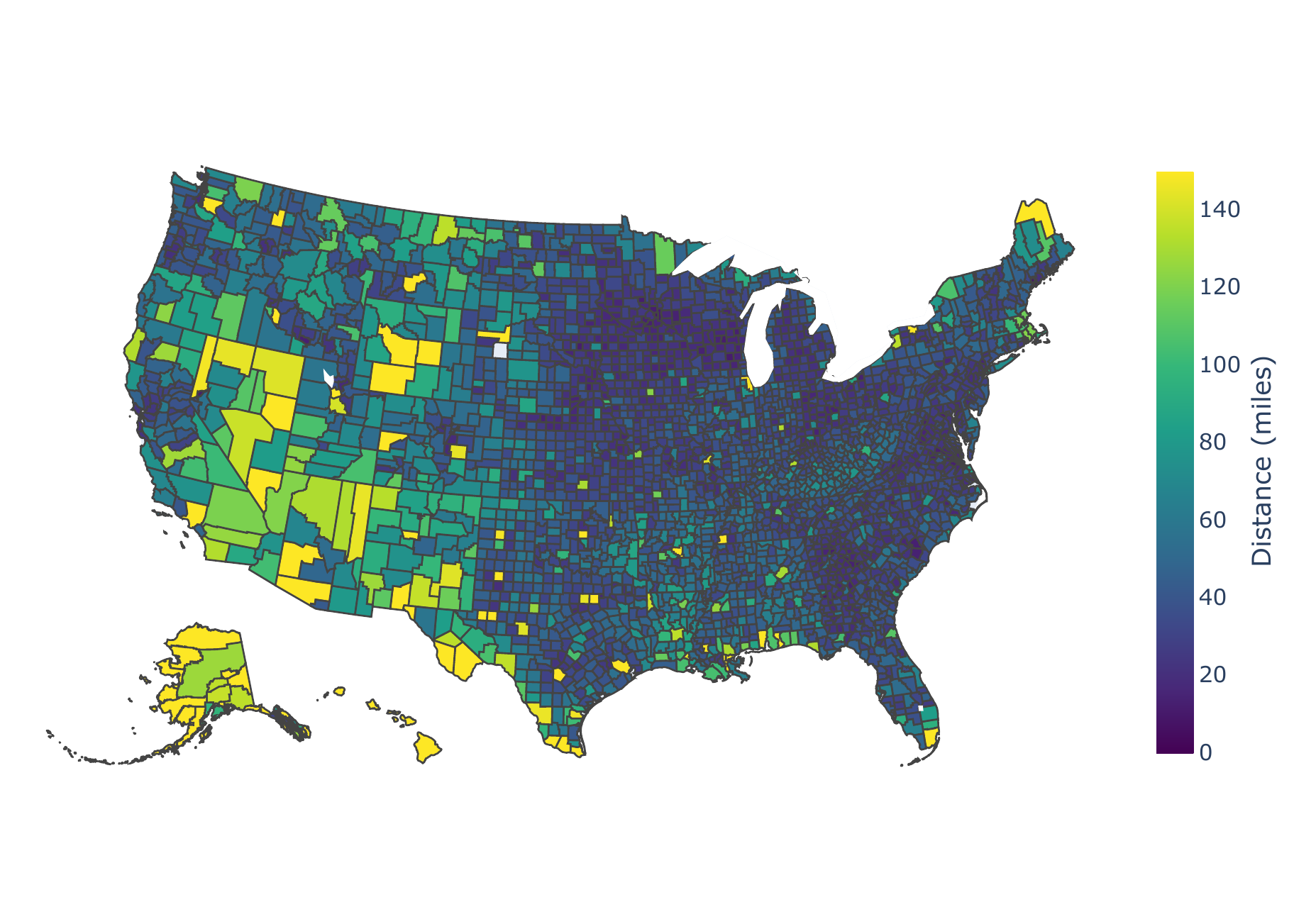}
    \caption{\textbf{The \acrfull{alf} dataset can be used to design and evaluate new metrics to measure access to assisted living for people in need.} This map shows the average distance of every \gls{alf} in the United States to people in need of assisted living (where need is defined in \Cref{sec:validation}).}
    \label{fig:need-map}
\end{figure}

\section{Need-based Metric for Access to Community-Based Care} 
\label{sec:need-metric}


Besides using the open dataset to validate socioeconomic and demographic characteristics of access to assisted living, we develop a need-based metric for directly assessing such access. Using data from the American Community Survey, we define someone's need for assisted living using the self-care, cognitive, ambulatory, and independent living difficulty variables. We designate an individual as being in need of an \glspl{alf} if they have at least two of \{independent living, cognitive, or ambulatory difficulty\}, \emph{and/or} if they have self-care difficulty. This definition of assisted living need is based on Medicaid waivers for reimbursement of the cost of assisted living: several states specify an individual must require assistance with at least two activities of daily living difficulties to qualify for an assisted living waiver \cite{mollica_state_2009}. The retrieval of American Community Survey data is described in Appendix \ref{sec:datasheets}.

To select a variable to assess access to assisted living, we use the location of an \gls{alf}. According to a report on how people choose nursing homes, the ``single most frequently cited factor in the selection of a facility was location'' \cite{shugarman_nursing_2006}, and we assume that this is the case for \glspl{alf} as well. We define the need-based metric per county as the average distance that an individual in need of assisted living must travel to reach a bed in an \gls{alf}. 

This metric is computed as the average distance from the centroid of each county to the set of facilities needed to satisfy the $n$ individuals in the county with assisted living need. This is accomplished using shapefiles of each county in the United States and determining their centroids. A k-d tree of all \glspl{alf} is created, with the value of the nodes being their capacity. For the 14\% of facilities without capacity information, their capacity is imputed as the mean of \glspl{alf} with known capacity. For each county, the nearest \glspl{alf} to the centroid of the county in the k-d tree are computed, with new facilities added to a running list until the sum of the capacities exceeds the assisted living need of the county. Since the coordinate system in the dataset is in latitude and longitude, the Haversine formula is used to approximate the distance between the centroid of a county and each of these facilities. The mean approximate distance to this subset of facilities is reported.

\paragraph{Conclusion.} Open data and reproducible, open source methods can lead to increased transparency and accountability of community-based care services such as \glspl{alf}. Open datasets such as what we present can also lead to better tools for people with which to decide where to live and receive care. 
We hope that this dataset will help researchers and policymakers use and develop machine learning methods to assess health disparity and expand access to community-based care facilities such as \glspl{alf} for the elderly and people with serious mental illness.

\begin{ack}
Thanks to Harry Reyes Nieva and Rajesh Ranganath for discussions, and to Ben Crewe and Boyd Bethel for submitting critical Freedom of Information Act requests.
%
\end{ack}

\printbibliography

\clearpage
\appendix
\input{sec_appendix}

\end{document}

%% file: preamble/preamble.tex
\usepackage[utf8]{inputenc} 
\usepackage[T1]{fontenc}    

\usepackage{url}            
\usepackage{booktabs}       
\usepackage{amsfonts}       
\usepackage{nicefrac}       
\usepackage{microtype}      
\usepackage[usenames,dvipsnames]{xcolor}

\colorlet{mylinkcolor}{violet}
\colorlet{mycitecolor}{Bittersweet} 
\colorlet{myurlcolor}{Aquamarine}
\usepackage[shortlabels]{enumitem}

\usepackage[colorinlistoftodos,prependcaption,textsize=tiny]{todonotes}

\usepackage{hyperref}       
\hypersetup{
  colorlinks=true,
   citecolor=Violet,
  linkcolor=Violet,
   urlcolor=MidnightBlue,
  linktoc=all
  }
\usepackage{graphicx}
\usepackage[labelfont=bf]{caption}
\usepackage[format=hang]{subcaption}
\usepackage{wrapfig}

\usepackage[acronym,nowarn]{glossaries}
\glsdisablehyper

\usepackage{cleveref}

\usepackage{booktabs}
\usepackage{array}

\usepackage{etoolbox, siunitx}
\robustify\bfseries

%% file: preamble/preamble_acronyms.tex
\newacronym{kl}{\textsc{kl}}{Kullback-Leibler}
\newacronym{elbo}{\textsc{elbo}}{evidence lower bound}
\newacronym{POPELBO}{pop-elbo}{\emph{population evidence lower bound}}
\newacronym{alf}{\textsc{alf}}{assisted living facility}
\newacronym{csv}{\textsc{csv}}{comma-separated value}
\newacronym{foia}{\textsc{foia}}{Freedom of Information Act}

%% file: preamble/preamble_biblatex.tex
\usepackage[%
minnames=1,
maxcitenames=2,
maxbibnames=99,
style=numeric, 
doi=false,
url=true,
giveninits=false, 
hyperref=true,
natbib=true,
backend=biber, 
uniquename=false,
uniquelist=false,
sorting=nyt
]{biblatex}%
\renewbibmacro{in:}{} 
\DeclareLanguageMapping{american}{american-apa}

\AtEveryBibitem{%
\ifentrytype{article}{
    \clearfield{url}%
    \clearfield{urldate}%
    \clearfield{eprint}
    \clearfield{series}
    \clearfield{volume}
}{}
\ifentrytype{book}{
    \clearfield{url}%
    \clearfield{urldate}%
}{}
\ifentrytype{collection}{
    \clearfield{url}%
    \clearfield{urldate}%
}{}
\ifentrytype{incollection}{
    \clearfield{url}%
    \clearfield{urldate}%
}{}
}

\AtEveryBibitem{
    \clearfield{pages}
    \clearfield{review}%
    \clearfield{series}
    \clearfield{volume}
    \clearfield{issue}
    \clearfield{number}
    \clearfield{pages}
    \clearfield{month}
    \clearfield{note}
    \clearfield{eprint}
    \clearfield{isbn}
    \clearfield{issn}
    \clearlist{location}
    \clearfield{series}
    \clearlist{publisher}
    \clearname{editor}
}{}

\DeclareFieldFormat{citehyperref}{%
  \DeclareFieldAlias{bibhyperref}{noformat}
  \bibhyperref{#1}}

\DeclareFieldFormat{textcitehyperref}{%
  \DeclareFieldAlias{bibhyperref}{noformat}
  \bibhyperref{%
    #1%
    \ifbool{cbx:parens}
      {\bibcloseparen\global\boolfalse{cbx:parens}}
      {}}}

\savebibmacro{cite}
\savebibmacro{textcite}

\renewbibmacro*{cite}{%
  \printtext[citehyperref]{%
    \restorebibmacro{cite}%
    \usebibmacro{cite}}}

\renewbibmacro*{textcite}{%
  \ifboolexpr{
    ( not test {\iffieldundef{prenote}} and
      test {\ifnumequal{\value{citecount}}{1}} )
    or
    ( not test {\iffieldundef{postnote}} and
      test {\ifnumequal{\value{citecount}}{\value{citetotal}}} )
  }
    {\DeclareFieldAlias{textcitehyperref}{noformat}}
    {}%
  \printtext[textcitehyperref]{%
    \restorebibmacro{textcite}%
    \usebibmacro{textcite}}}

%% file: table_replication.tex
\begin{table}[tb]
\centering
\resizebox{1.0\textwidth}{!}{%
\begin{tabular}{l
S[table-format=3.1] @{${}\pm{}$} S[table-format=2.1]
S[table-format=3.1] @{${}\pm{}$} S[table-format=2.1]
S[table-format=3.1] @{${}\pm{}$} S[table-format=2.1]
S[table-format=3.1] @{${}\pm{}$} S[table-format=2.1]
S[table-format=3.1] @{${}\pm{}$} S[table-format=3.1]}
\toprule
 & \multicolumn{10}{c}{Assisted Living Facility Penetration} \\
County Characteristic & \multicolumn{2}{c}{No Facilities} & \multicolumn{2}{c}{Quartile 1} & \multicolumn{2}{c}{Quartile 2} & \multicolumn{2}{c}{Quartile 3} & \multicolumn{2}{c}{Quartile 4} \\
\midrule
\# facilities per 1000 people $\geq 65$ & \multicolumn{2}{c}{} &  \multicolumn{2}{c}{0.06--12.40} & \multicolumn{2}{c}{12.40--21.38} & \multicolumn{2}{c}{21.38--32.34} & \multicolumn{2}{c}{$\geq$32.34} \\
Number of Counties & \multicolumn{2}{c}{1204} & \multicolumn{2}{c}{507} & \multicolumn{2}{c}{506} & \multicolumn{2}{c}{506} & \multicolumn{2}{c}{507} \\
Median age & 42.1 & 5.4 & 42.4 & 5.1 & 41.2 & 5.2 & 40.8 & 5.3 & 39.9 & 5.3 \\
Percent of population, age $\geq 65$  & 19.3&	4.7&	19.4&	4.3&	18.7&	4.5&	18.3&	4.9&	17.6&	4.3 \\
Percent of population, age $\geq 85$ & 2.3&	1.0&	2.2	&0.7	&2.1	&0.7&	2.2	&1.0&	2.3&	1.0 \\
Less than high school  & 14.7&	8.0&	14.1&	6.0&	13.5&	5.2	&12.0&	5.0&	10.6&	4.8 \\
College education or higher  & 19.8	&8.2	&19.8&	8.6&	21.1&	8.0&	24.6&	9.5	&27.5&	11.6 \\
Median household income (\$1000s) & 48.7&	14.7&	51.3&	13.2&	52.2&	12.6&	56.6&	14.7&	59.6	&16.4 \\
Unemployment rate  & 6.4&	2.6&	7.1&	2.1&	7.3&	2.2	&6.9	&2.0&	6.5	&2.1 \\
Poverty rate &17.8&	10.8&	15.9&	5.9	&15.6&	5.5	&14.5&	5.5	&13.2&	5.2 \\
Home ownership rate & 73.0&	8.3	&72.6&	7.8	&71.3&	7.5	&70.1&	7.9	&68.8&	9.2 \\
Median owned home value (\$1000s) & 127.0&	71.8&	149.4&	94.5&	148.7	&69.4	&172.6&	92.1&	199.2&	135.6 \\
Percent white population  & 83.4&	18.5&	82.0&	17.3&	81.0&	17.3&	81.6&	16.2&	83.1&	14.7 \\
Percent black population & 7.7&	14.5&	10.7&	16.0	&11.0	&15.1&	10.2&	14.1&	8.0&	12.8\\
Percent hispanic population  & 15.1&	26.8&	10.4&	15.9&	9.2	&12.8&	9.2	&10.8&	9.3	&10.9 \\
Ratio men to women (100s) & 101.8&	13.1&	101.4&	10.8&	100.1&	10.7&	98.6&	7.9	&99.8&	9.4 \\
\bottomrule
\end{tabular}
}
\vspace{1ex}
\caption{\textbf{The \acrfull{alf} dataset presented here can be used to replicate existing work by \citet{cornell_developments_2020} on access to community-based care, by computing county traits by assisted living penetration per quartile.} County characteristics and statistics are drawn from the 2015--2019 American Community Survey data~\cite{noauthor_american_nodate}, besides the 2020 unemployment rate~\cite{noauthor_local_nodate}.}
\label{table:replication}
\end{table}

%% file: sec_appendix.tex
\section{Description of Variables in Assisted Living Dataset}
\label{sec:variable-description}

\input{table_alf_variables}

The table contains one row for each variable in the dataset. A description of the variable is given. The Percent Filled column describes what percent of \glspl{alf} in the dataset contain the respective variable.

\section{Datasheets for Datasets}
\label{sec:datasheets}

We use the dataset documentation framework from \citet{gebru_datasheets_2020}.

\subsection{Motivation}

\textbf{For what purpose was the dataset created? Was there a specific task in mind? Was there a specific gap that needed to be filled? Please provide a description.}

A key part of the Biden administration's \$2 trillion infrastructure plan is \$400 billion allocated to home and community-based care in the United States \cite{noauthor_fact_2021}. However, basic questions about where, how, and to whom these monies and resources should be allocated remain, and there is no public dataset with which to answer these questions for policymakers, researchers, and the public. Machine learning methods to assess cost, health disparities, and quality of assisted living facilities will be key to planning and maintenance of this public health infrastructure. This dataset was created in support of these aims.

\textbf{Who created the dataset (e.g., which team, research group) and on
behalf of which entity (e.g., company, institution, organization)?}

Anton Stengel and Jaan Altosaar created the dataset on behalf of research activities in the Elhadad Lab at Columbia University.
\clearpage
\textbf{Who funded the creation of the dataset? If there is an associated
grant, please provide the name of the grantor and the grant name and
number.}

Anton Stengel was funded by Princeton University Center for Career Development.

\textbf{Any other comments?} [N/A]

\subsection{Composition}

\textbf{What do the instances that comprise the dataset represent (e.g.,
documents, photos, people, countries)? Are there multiple types of
instances (e.g., movies, users, and ratings; people and interactions between them; nodes and edges)? Please provide a description.}

The instances represent assisted living facilities in the United States, which are licensed homes for the elderly and people with serious mental illness to receive food, shelter, care, help with activities of daily living and other necessities.

\textbf{How many instances are there in total (of each type, if appropriate)?}

There are \numberofalfs~assisted living facilities in the dataset.

\textbf{Does the dataset contain all possible instances or is it a sample
(not necessarily random) of instances from a larger set? If the
dataset is a sample, then what is the larger set? Is the sample representative of the larger set (e.g., geographic coverage)? If so, please describe how
this representativeness was validated/verified. If it is not representative
of the larger set, please describe why not (e.g., to cover a more diverse
range of instances, because instances were withheld or unavailable).}

The dataset contains all possible instances of licensed assisted living facilities in the United States as of 3/24/21. Data was collected between 6/24/21 and 8/24/21. The conservative date is three months prior to the earliest access point because some states' licensing agencies update their public databases on a quarterly basis.

\textbf{What data does each instance consist of? “Raw” data (e.g., unprocessed text or images) or features? In either case, please provide a description.}

Each instance consists of raw data and cross-linked or pre-processed variables. The variables associated with every assisted living facility are in \Cref{table:alf-variables}.

\textbf{Is there a label or target associated with each instance? If so, please
provide a description.}

No.

\textbf{Is any information missing from individual instances? If so, please
provide a description, explaining why this information is missing (e.g.,
because it was unavailable). This does not include intentionally removed
information, but might include, e.g., redacted text.}

Yes. States' licensing agencies provided different levels of detail about individual instances. The missing data was not filled in, besides for county information and facility type, which were manually inputted when empty.

\textbf{Are relationships between individual instances made explicit
(e.g., users’ movie ratings, social network links)? If so, please describe how these relationships are made explicit.}

Yes. Assisted living facilities within the same state are licensed by the same licensing agency.

\textbf{Are there recommended data splits (e.g., training, development/validation,
testing)? If so, please provide a description of these splits, explaining
the rationale behind them.}

No.

\textbf{Are there any errors, sources of noise, or redundancies in the
dataset? If so, please provide a description.}

Optical character recognition, HTML parsing, and manual transcription was required for data from some states in the dataset, which may induce some errors. A small amount of noise from optical character recognition or manual error is possible for instances in these states.

Some states' directories of licensed facilities seemed to contain occasional misspellings, formatting issues, duplicate facilities, and closed facilities. This could lead to a small amount of error in the dataset.

\textbf{Is the dataset self-contained, or does it link to or otherwise rely on
external resources (e.g., websites, tweets, other datasets)? If it links
to or relies on external resources, a) are there guarantees that they will exist, and remain constant, over time?; b) are there official archival versions
of the complete dataset (i.e., including the external resources as they existed at the time the dataset was created); c) are there any restrictions
(e.g., licenses, fees) associated with any of the external resources that
might apply to a future user? Please provide descriptions of all external
resources and any restrictions associated with them, as well as links or
other access points, as appropriate.}

The dataset is self-contained.

\textbf{Does the dataset contain data that might be considered confidential (e.g., data that is protected by legal privilege or by doctor-
patient confidentiality, data that includes the content of individuals’ non-public communications)? If so, please provide a description.}

No.

\textbf{Does the dataset contain data that, if viewed directly, might be offensive, insulting, threatening, or might otherwise cause anxiety?
If so, please describe why.}

No.

\textbf{Does the dataset relate to people? If not, you may skip the remaining
questions in this section.}

No, the dataset does not relate to individual people. It is a database of assisted living facilities, and to assess assisted living need, it includes aggregate information about people at a population and county level.

\textbf{Does the dataset identify any subpopulations (e.g., by age, gender)? If so, please describe how these subpopulations are identified and
provide a description of their respective distributions within the dataset.}

The dataset identifies subpopulations of the United States by some demographic characteristics at the county level. Their distributions are given in \Cref{table:replication}.

\textbf{Is it possible to identify individuals (i.e., one or more natural persons), either directly or indirectly (i.e., in combination with other
data) from the dataset? If so, please describe how.}

No. Public use microdata areas defined by the Bureau of the Census contain no fewer than 100,000 people each, and we employ statistics at the microdata level \cite{bureau_sample_nodate}.

\textbf{Does the dataset contain data that might be considered sensitive
in any way (e.g., data that reveals racial or ethnic origins, sexual
orientations, religious beliefs, political opinions or union memberships, or locations; financial or health data; biometric or genetic data; forms of government identification, such as social security numbers; criminal history)? If so, please provide a description.}

No. The race, age, gender variables are at a de-identified microdata area level.

\textbf{Any other comments?} [N/A]

\subsection{Collection Process}

\textbf{How was the data associated with each instance acquired? Was
the data directly observable (e.g., raw text, movie ratings), reported by
subjects (e.g., survey responses), or indirectly inferred/derived from other
data (e.g., part-of-speech tags, model-based guesses for age or language)?
If data was reported by subjects or indirectly inferred/derived from other
data, was the data validated/verified? If so, please describe how.}

See \ref{sec:building-dataset} for information on how the \gls{alf} data was collected.

Here we describe how the need-based metric data and some \gls{alf} location data was collected.

To compute the need-based metric, The Census's Microdata Access Tool (MDAT) was queried to get Public Use Microdata Samples (PUMS) of around three million individuals in the United States in 2019 \citep{noauthor_american_nodate}. Each PUMS instance consists of an anonymized individual with variables for self-care, cognitive, ambulatory, and independent living difficulty, as well as the specific Public Use Microdata Area (PUMA) within which the individual is located. PUMAs are geographic areas that partition each state into areas containing at least 100,000 people each. Whether each individual has assisted living need was computed as a function of the four variable disabilities \ref{sec:need-metric}. The number of individuals with assisted living need in each PUMA was counted. Then the count per PUMA was converted to count per county using the Missouri Census Data Center's PUMA-county equivalence file, which provides the overlapping percentage of population between every physically overlapping PUMA and county in the United States \citep{noauthor_geocorr_nodate}. We used a weighted sum of the overlapping populations to estimate the statistic per county.

Location data was then added to each \gls{alf}. Google Maps Geocoding API was queried to get latitude and longitude information and county name information for each \gls{alf} with a corresponding address. The Federal Communications Commission Area and Census Block API was then queried to get the county Federal Information Processing Standards code for each \gls{alf}, which are a five-digit code which uniquely identifies counties and county equivalents in the United States. These codes are useful for linking the assisted living data to county statistics.

\textbf{What mechanisms or procedures were used to collect the data
(e.g., hardware apparatus or sensor, manual human curation, software program, software API)? How were these mechanisms or procedures validated?}

Manual human curation and software programs. These mechanisms and procedures were validated by cross-referencing the datasets with states' assisted living facilities, and comparing the total number of facilities to previous work on similar data \cite{cornell_developments_2020}.

\textbf{If the dataset is a sample from a larger set, what was the sampling
strategy (e.g., deterministic, probabilistic with specific sampling
probabilities)?} [N/A]

\textbf{Who was involved in the data collection process (e.g., students,
crowdworkers, contractors) and how were they compensated (e.g.,
how much were crowdworkers paid)?} 

The lead author led the data collection process and was compensated \$600/week for the duration of the summer.

\textbf{Over what timeframe was the data collected? Does this timeframe
match the creation timeframe of the data associated with the instances
(e.g., recent crawl of old news articles)? If not, please describe the timeframe in which the data associated with the instances was created.}

Data was collected between 6/24/21 and 8/24/21. This broadly matches the creation timeframe of the data associated with the instances; some states update their directories as new facilities are licensed, while other states only update their directories on a quarterly basis.

\textbf{Were any ethical review processes conducted (e.g., by an institutional review board)? If so, please provide a description of these review
processes, including the outcomes, as well as a link or other access point
to any supporting documentation.}

No.

\textbf{Does the dataset relate to people? If not, you may skip the remainder
of the questions in this section.}

Yes.

\textbf{Did you collect the data from the individuals in question directly,
or obtain it via third parties or other sources (e.g., websites)?}

Aggregate population- and county-level statistics were obtained from the Bureau of the Census. 

\textbf{Were the individuals in question notified about the data collection? If so, please describe (or show with screenshots or other information) how notice was provided, and provide a link or other access point
to, or otherwise reproduce, the exact language of the notification itself.} [N/A]

\textbf{Did the individuals in question consent to the collection and use
of their data? If so, please describe (or show with screenshots or other
information) how consent was requested and provided, and provide a
link or other access point to, or otherwise reproduce, the exact language
to which the individuals consented.} [N/A]

\textbf{If consent was obtained, were the consenting individuals provided with a mechanism to revoke their consent in the future or
for certain uses? If so, please provide a description, as well as a link or
other access point to the mechanism (if appropriate).} [N/A]

\textbf{Has an analysis of the potential impact of the dataset and its use
on data subjects (e.g., a data protection impact analysis) been conducted? If so, please provide a description of this analysis, including
the outcomes, as well as a link or other access point to any supporting
documentation.} [N/A]

\textbf{Any other comments?} [N/A]

\subsection{Preprocessing/cleaning/labeling}
\label{sec:data-cleaning}

\textbf{Was any preprocessing/cleaning/labeling of the data done (e.g.,
discretization or bucketing, tokenization, part-of-speech tagging,
SIFT feature extraction, removal of instances, processing of missing values)? If so, please provide a description. If not, you may skip the
remainder of the questions in this section.}

After collecting data for all 48 states, data cleaning was carried out. First, the facility name and address information were parsed into well-formatted columns and capitalization was standardized. Next, duplicate facilities were removed.
Duplicates were identified as facilities that have identical names and license numbers or identical names and facility identifiers. For states where there was no license number or facility identifiers available, we considered facilities to be duplicates if they had identical names and addresses.

\textbf{Was the “raw” data saved in addition to the preprocessed/cleaned/labeled
data (e.g., to support unanticipated future uses)? If so, please provide a link or other access point to the “raw” data.}

Yes. It is in the linked repository.

\textbf{Is the software used to preprocess/clean/label the instances available? If so, please provide a link or other access point.}

Yes. It is in the linked repository.

\textbf{Any other comments?} [N/A]

\subsection{Uses}

\textbf{Has the dataset been used for any tasks already? If so, please provide
a description.} 

No.

\textbf{Is there a repository that links to any or all papers or systems that
use the dataset? If so, please provide a link or other access point.}

Yes. The link will be made available upon publication.

\textbf{What (other) tasks could the dataset be used for?}

Besides developing metrics to assess assisted living need and availability as in the present work, machine learning methods could be used to assess health disparity, inequality in access to assisted living, and various other tasks.

\textbf{Is there anything about the composition of the dataset or the way
it was collected and preprocessed/cleaned/labeled that might impact future uses? For example, is there anything that a future user
might need to know to avoid uses that could result in unfair treatment
of individuals or groups (e.g., stereotyping, quality of service issues) or
other undesirable harms (e.g., financial harms, legal risks) If so, please
provide a description. Is there anything a future user could do to mitigate
these undesirable harms?}

No.
\clearpage
\textbf{Are there tasks for which the dataset should not be used? If so,
please provide a description.}

The dataset is de-identified, and in accordance with data privacy practices should not be used in an attempt to identify individual people living at the assisted living facility locations. A similar dataset already exists for nursing homes provided by the Centers for Medicare \& Medicaid Services \citep{noauthor_provider_2021}, where similar considerations apply. We acknowledge that the assisted living facility dataset elevates the addresses or prominance of some smaller facilities into the public eye, but that this information is already a matter of public record through state licensing agencies. The dataset, as-is, should not be used to assess the quality or appropriateness of a facility for an individual. There is no assessment of quality or recommendation of whether the facilities are appropriate for any individual in need of assisted living.

\textbf{Any other comments?} [N/A]

\subsection{Distribution}

\textbf{Will the dataset be distributed to third parties outside of the entity (e.g., company, institution, organization) on behalf of which
the dataset was created? If so, please provide a description.}

No.

\textbf{How will the dataset will be distributed (e.g., tarball on website,
API, GitHub)? Does the dataset have a digital object identifier (DOI)?}

On GitHub and on the website. There is no digital object identifier yet.

\textbf{When will the dataset be distributed?}

Upon publication.

\textbf{Will the dataset be distributed under a copyright or other intellectual property (IP) license, and/or under applicable terms of use
(ToU)? If so, please describe this license and/or ToU, and provide a link
or other access point to, or otherwise reproduce, any relevant licensing
terms or ToU, as well as any fees associated with these restrictions.}

No.

\textbf{Have any third parties imposed IP-based or other restrictions on
the data associated with the instances? If so, please describe these
restrictions, and provide a link or other access point to, or otherwise
reproduce, any relevant licensing terms, as well as any fees associated
with these restrictions.}

No.

\textbf{Do any export controls or other regulatory restrictions apply to
the dataset or to individual instances? If so, please describe these
restrictions, and provide a link or other access point to, or otherwise
reproduce, any supporting documentation.}

Regulatory restrictions apply to individual assisted living facilities, and vary state-by-state.

\textbf{Any other comments?} [N/A]

\subsection{Maintenance}

\textbf{Who is supporting/hosting/maintaining the dataset?}

\url{https://onefact.org}

\textbf{How can the owner/curator/manager of the dataset be contacted
(e.g., email address)?}

Anton Stengel and Jaan Altosaar can be contacted at \href{mailto:astengel@princeton.edu}{astengel@princeton.edu} and \href{mailto:j@jaan.io}{j@jaan.io}.

\textbf{Is there an erratum? If so, please provide a link or other access point.}

The erratum will be kept updated on the GitHub page, \url{https://github.com/onefact/assisted-living}.
\clearpage
\textbf{Will the dataset be updated (e.g., to correct labeling errors, add
new instances, delete instances)? If so, please describe how often, by
whom, and how updates will be communicated to users (e.g., mailing list,
GitHub)?}

The dataset will be updated through pull requests submitted via GitHub at anytime by anyone. Any dataset updates will be reviewed and merged by team members.

\textbf{If the dataset relates to people, are there applicable limits on the
retention of the data associated with the instances (e.g., were individuals in question told that their data would be retained for a
fixed period of time and then deleted)? If so, please describe these
limits and explain how they will be enforced.} [N/A]

\textbf{Will older versions of the dataset continue to be supported/hosted/maintained?
If so, please describe how. If not, please describe how its obsolescence
will be communicated to users.}

Yes. Older versions of the dataset will continue to be hosted on GitHub.

\textbf{If others want to extend/augment/build on/contribute to the
dataset, is there a mechanism for them to do so? If so, please
provide a description. Will these contributions be validated/verified?
If so, please describe how. If not, why not? Is there a process for communicating/distributing these contributions to other users? If so, please
provide a description.}

Yes. This paper describes the process for how the dataset was collected, and anyone can build on this dataset by following the outlined steps. Contributions will be reviewed as pull requests to the GitHub repository.

\textbf{Any other comments?} [N/A]

%% file: table_alf_variables.tex
\begin{table}[h!]
\centering
\resizebox{1.0\textwidth}{!}{
\begin{tabular}{llc}
\toprule
Variable & Description & Percent Filled
\\
\midrule
Facility Name	&	Name of the facility	&	100\%	\\
Facility ID	&	Facility identification number	&	65\%	\\
License Number	&	Facility license number	&	48\%	\\
Address	&	Primary physical address of the facility	&	100\%	\\
City	&	City of the facility	&	98\%	\\
State	&	State of the facility	&	100\%	\\
Zip Code	&	Zip code of the facility	&	97\%	\\
County	&	County of the facility	&	100\%	\\
County FIPS	&	County identification code	&	100\%	\\
Latitude	&	Latitude of the facility	&	100\%	\\
Longitude	&	Longitude of the facility	&	100\%	\\
Facility Type Primary	&	Primary licensing type of the facility	&	100\%	\\
Facility Type Secondary	&	Secondary licensing type of the facility	&	41\%	\\
Capacity	&	Total capacity of the facility (number of beds)	&	86\%	\\
Ownership Type	&	Ownership structure of the facility	&	27\%	\\
Licensee	&	The license holder of the facility	&	48\%	\\
Phone Number	&	Phone number associated with facility	&	98\%	\\
Email Address	&	Email address associated with facility	&	35\%	\\
Date Accessed	&	Date facility information retrieved from state licensing agency	&	100\%	\\
Total County AL Need	&	Computed need-based metric for county of facility	&	100\%	\\
County Percent of Population 65 or Older	&	Retrieved from 2015-2019 ACS data	&	100\%	\\
County Median Age	&	Retrieved from 2015-2019 ACS data	&	100\%	\\
County Homeownership Rate	&	Retrieved from 2015-2019 ACS data	&	100\%	\\
County College Education or Higher Rate	&	Retrieved from 2015-2019 ACS data	&	100\%	\\
County Percent Black Population	&	Retrieved from 2015-2019 ACS data	&	100\%	\\
County Median Home Value of Owned Homes	&	Retrieved from 2015-2019 ACS data	&	100\%	\\
County Percent Hispanic Population	&	Retrieved from 2015-2019 ACS data	&	100\%	\\
County Percent of Population 85 or Older	&	Retrieved from 2015-2019 ACS data	&	100\%	\\
County Meidan Household Income	&	Retrieved from 2015-2019 ACS data	&	100\%	\\
County Unemployment Rate	&	Retrieved from 2020 ACS data	&	100\%	\\
County Less Than High School Diploma Rate	&	Retrieved from 2015-2019 ACS data	&	100\%	\\
County Percent Whilte Population	&	Retrieved from 2015-2019 ACS data	&	100\%	\\
County Poverty Rate	&	Retrieved from 2015-2019 ACS data	&	100\%	\\
\bottomrule
\end{tabular}
}
\vspace{1ex}
\caption{\textbf{Variables associated with every \acrlong{alf} in this open dataset.}}
\label{table:alf-variables}
\end{table}